\documentclass[12pt]{article}
\usepackage[cp1251]{inputenc}
\usepackage[english]{babel}
\usepackage{amssymb}
\usepackage{graphicx}
\graphicspath{}
\DeclareGraphicsExtensions{.pdf,.png,.jpg}
\begin{document}
\large

\begin{center}
{\Large Asymptotic step-like solutions to the singularly perturbed Burgers' equation}
\end{center}
\noindent

Samoilenko Valerii \\
Department of Mathematical Physics, Taras Shevchenko National University of Kyiv

\vskip5mm

Samoilenko Yuliia\\
Institute Camille Jordan, Claude Bernard Lyon 1 University

\vskip5mm

Zappale Elvira\\
Department of Basic and Applied Sciences for Engineering, Sapienza - University of Rome

\begin{abstract}

The paper deals with a problem of asymptotic step-like solutions to the Burgers' equation with variable coefficients and a small parameter.
By means of the non-linear WKB method, the algorithm of constructing these asymptotic solutions is proposed and statements on justification of the algorithm are proved.

The obtained results are illustrated by an example, for which the first asymptotic step-like approximation is explicitly found.
The asymptotic solution is global, and has a form of the shock wave type function. There are also given graphs of these approximate solutions for certain numerical parameters.
\\[2mm]
MSC Class: 35Q35; 35B25; 35C20; 76D33; 76M45
\\[2mm]

\end{abstract}

\section{Introduction} The paper deals with constructing asymptotic solutions to the singularly perturbed Burgers' equation with variable coefficients of the following form
\begin{equation} \label{Burgers_VarCoef}
\label{Burgers_v} \varepsilon u_{xx} = a(x, t, \varepsilon) u_t +
b(x, t, \varepsilon) u u_x, \quad n \in {\mathbf N},
\end{equation}
where the functions $ a(x, t, \varepsilon) $, $ b(x, t, \varepsilon) $, $ (x,
t) \in {\mathbf R} \times [0; T] $, are written as asymptotic (according to Poincar\'e) series
\begin{equation} \label{coeff}
a(x, t, \varepsilon) = \sum\limits_{k=0}^N \varepsilon^k a_k(x, t) +
O(\varepsilon^{N+1}) , \ b(x, t, \varepsilon) = \sum\limits_{k=0}^N
\varepsilon^k b_k(x, t) + O(\varepsilon^{N+1}).
\end{equation}
Here $ a_0(x, t) \, b_0(x, t)\not= 0 $ for all $ (x, t) \in
{\mathbf R} \times [0; T] $; $ \varepsilon $ is a small parameter.

Equation (\ref{Burgers_VarCoef}) is a mathematical model of wave processes in medium with small viscosity and variable characteristics and it is a generalization of the Burgers'  equation with constant coefficients
\begin{equation} \label{Burgers}
u_t + uu_x = \nu u_{xx}, 
\end{equation}
where $ u=u(x,t) $ is velocity of liquid moving and $ \nu $ is viscosity of liquid.

The equation is used in hydrodynamics, acoustic, gas-dynamic and other sections of applied mathematics while mathematical description of wave processes with viscosity $ \nu >0 $.

Equation (\ref{Burgers}) has been known since 1906 (see \cite{Forsyth}, \cite{Bateman}) and has attracted the attention of researchers, because in one-dimensional case the equation is a simple form of Navier-Stokes equation, one of the most important model of hydrodynamics \cite{Sedov}.

Recall that the Navier-Stokes system was obtained in the first part of the XIX-th century, first by A.~Navier, then by O.~Cauchy, S.~Poisson, A.~Saint-Venant and G.~Stokes and describes moving of a viscous (in the general case, incompressible) liquid (gas) and is based on Newton's laws.

The Navier-Stokes system is non-linear that is essentially complicate studying its solutions which are sensitive to perturbation of the coefficients as well as to the initial (or/and the boundary) conditions.


At the end of the 30-th years of the past century J.~M.~Burgers studied equation (\ref{Burgers}) as the simplest model unifying typical non-linearity and viscosity which describes phenomenon of hydrodynamic turbulence \cite{Burgers, Burgers1, Burgers2}.
If the viscosity is absent (for $\nu =0$ ) then equation (\ref{Burgers}) is transmission equation
\begin{equation} \label{Hopf}
u_t + uu_x=0,
\end{equation}
that is known as the Hopf equation \cite{Hopf}.

For solution to the equation so-called \textit{gradient catastrophe} takes place \cite{ZasSagdeev}. The phenomenon is connected with overturning (destroying) waves. It is deduced from the formula for the solution to the Cauchy problem for equation (\ref{Hopf}) with initial condition $ u(x,0)=f(x), $ $ x\in {\mathbf R}, $ that is implicitly written as $ u(x,t)=f(x - t u(x,t))) $. In particular, from the note it can be concluded that if the initial function $ f(x) $ is quickly decreasing, then the solution to the corresponding Cauchy problem for equation (\ref{Hopf}) exists only on finite time interval \cite{SamYu}. The properties are obviously consequence of presence of non-linear term $ u u_x$. Even a small perturbation of the form $ \nu u_{xx} $, where $ \nu >0 $ in equation (\ref{Hopf}), significantly affect the qualitative properties of its solutions.

Model (\ref{Burgers}) is one of the simplest describing non-linear effects, in particular, appearance and evolution of shock waves. Many qualitative properties of Burgers' equation (\ref{Burgers}) can be obtained from formulas for its exact solutions, which are found by reducing nonlinear equation (\ref{Burgers}) to the heat linear equation using the substitution proposed by E.~Hopf \cite{Hopf} and J.~Cole \cite{Cole} independently.
In such a way they got formulas for exact solutions to the Burgers' equation, that let to analyze process of appearance shock waves for non-linear environment with a small viscosity from the initial conditions. In particular, because of viscosity there is not overturning waves, but presence of the dissipation induces appearance of the so-called stationary waves prolonging with constant velocity \cite{ZasSagdeev,
Ablowitz}. These stationary solutions to equation (\ref{Burgers})
are written as
\begin{equation} \label{sol_Burgers_pert}
u(x,t) = a\left(1 - \tanh{ \left( \frac{a}{2} \, \frac{x-at}{\nu} + \varphi_0 \right)} \right) ,
\end{equation}
where the real constants $a$, $ \varphi_0 $ are arbitrary.

In the case of \textit{disappearing viscosity}, i.e. if $ \nu \to 0 $,
then the limit value of the solution to equation (\ref{sol_Burgers_pert}) is discontinuous step-like function, because
\begin{equation}\label{sol_Burgers_pert_1}
\lim\limits_{\nu \to \, 0} u(x,t) = \left\{\begin{array}{cc}
       0,   & x - at > 0, \\
       2 a, & x - at < 0. \\
\end{array}
\right.
\end{equation}

Therefore, there is naturally appearanced a problem on constructing approximate solutions to equation  (\ref{Burgers_VarCoef}) with a small parameter $ \varepsilon $ that analogously to the case of equation (\ref{Burgers}) in their asymptotic behavior (as $ \nu \to 0 $) are similar to solutions of the form (\ref{sol_Burgers_pert}).

This problem is considered in the paper, where using the WKB-technique for equation (\ref{Burgers_VarCoef}) there is constructed the asymptotic solution that is a discontinuous step-like function in limits case (as $ \varepsilon \to 0 $) \, \cite{Maslov, SamKdF, SamBBM}. Such solution is called an asymptotic step-like solution.

In the paper, we propose a general algorithm of constructing these solutions, find the first approximation for the asymptotic step-like solution and prove the theorems on the accuracy with which the main term and the first approximation satisfy equation (\ref{Burgers_VarCoef}).
These results are demonstrated on an example.

\section{Preliminary notes and definitions}

According to the concept of constructing asymptotic solutions to the singularly perturbed differential equations of hydrodynamics with variable coefficients the searched solution can be written as asymptotic series in the following form \cite{SamKdF, SamBBM}:
\begin{equation} \label{as_sol}
u(x, t, \varepsilon) = \sum\limits_{j=0}^{\infty} \varepsilon^j
\left[ u_j(x, t) + V_j (x, t, \tau) \right] , \quad \tau = \frac{x - \varphi(t)}{\varepsilon},
\end{equation}
where $ u_j(x, t) $, $ V_j(x, t, \tau) $, $ j = 0,1, \ldots \, $, $ \varphi = \varphi(t) $ are infinitely differentiable functions that are defined while constructing the asymptotic solution.

Here the function $ \varphi = \varphi (t) $ defines the discontinuity curve \cite{Maslov, SamKdF}.

The function $ U(x, t, \varepsilon) = \sum\limits_{j=0}^{\infty} \varepsilon^j u_j(x,t)$ is called the regular part of asymptotic (\ref{as_sol}), and the function
$ V(x, t, \tau , \varepsilon) = \sum\limits_{j=0}^{\infty} \varepsilon^j \, V_j (x, t, \tau) $ is called the singular part of the asymptotic. The regular part is a background function, while its  singular part describes qualitative properties of searched asymptotic solution (\ref{as_sol}).

According to properties of functions (\ref{sol_Burgers_pert}) the terms of the singular part of the asymptotics are searched as elements of certain functional space $ G $, that is defined as follows \cite{SamKdF, SamBBM}.

The space $ G = G({\bf R} \times [0; T] \times {\bf R}) $ contains functions $ f = f(x, t, \tau) \in C^{\, \infty} ({\bf R} \times [0; T] \times {\bf R}) $, such that the following conditions are fulfilled: \\
$1^0$. the relation
$$
\lim\limits_{\tau \to + \infty} \tau^n \frac{\partial
\,^{p+q+r}}{\partial x^p \, \partial t^q \,
\partial \tau^r} \, f (x, t, \tau) = 0, \quad (x, t) \in K,
$$
is satisfied for all non-negative integers $ n $, $ p $, $ q $, $ r $ uniformly with respect to $ (x, t) $
on any compact $ K \subset {\bf R} \times [0; T] $;\\
$2^0$. there exists an infinitely differentiable function $ f^-(x, t) $ such that
$$
\lim\limits_{\tau \to - \infty} \tau^{n} \frac{\partial
\,^{p+q+r}}{\partial x^p \, \partial t^q \,
\partial \tau^r} \, \left( f (x, t,
\tau) - f^{-}(x, t)\right) = 0, \quad (x, t) \in K .
$$

Remark that according to notation $ \displaystyle \tau = \frac{x-\varphi (t)}{\varepsilon}$  with $ \varphi (t) = at $, function (\ref{sol_Burgers_pert}) belongs to the space $G$. From property (\ref{sol_Burgers_pert_1}) for function (\ref{sol_Burgers_pert}) it follows the appropriateness of using notice of discontinuity curve, which is defined as
$ \Gamma = \{(x,t)\in {\mathbf R}\times [0; T]: x = \varphi(t)\} $ for solution (\ref{as_sol}).

By $ G_0 = G_0 ({\bf R} \times [0; T] \times {\bf R}) \subset G $ we denote the space of such functions $ f = f(x,t, \tau) $ that the condition
$$
\lim_{\tau \to - \infty} f (x, t, \tau) = 0
$$
holds uniformly with respect to $ (x, t) \in K $ for any compact $ K \subset {\bf R} \times [0; T] $.

\section{Algorithm of constructing the asymptotic solution}

While finding the terms of asymptotic solution ({\ref{as_sol}}) we use general approaches and ideas of asymptotic analysis as well as properties of the space $ G $.
For asymptotic series ({\ref{as_sol}}) firstly we determine the terms of the regular part and then we find the terms of the singular part of the asymptotics as well as the function $ \varphi = \varphi(t) $ defining the discontinuity curve.

Equations for the regular terms of the asymptotics have the following form
\begin{equation} \label{reg_part_0}
a_0(x, t) \frac{\partial u_0}{\partial t} + b_0(x, t) u_0 \frac{\partial u_0}{\partial x} = 0,
\end{equation}
\begin{equation} \label{reg_part_1}
a_0(x, t) \frac{\partial u_j}{\partial t} + b_0(x, t) \, \frac{\partial}{\partial x} \left( u_0 \, u_j \right)  = f_j(x, t),
\end{equation}
and are obtained after substituting solution ({\ref{as_sol}}) in (\ref{Burgers_v}) using the property $1^0$ for $ V_j (x, t,
\tau) \in G $, $ j=0,1, \ldots \, $\,.

The functions $ f_j(x, t), $ $ j=0,1, \ldots \, $, \, in \, (\ref{reg_part_1}) are found recurrently. For example,
$$
f_1(x,t) = \frac{\partial^2 u_0}{\partial x^2} - a_1(x, t) \frac{\partial u_0}{\partial t} - b_1(x, t) u_0 \frac{\partial u_0}{\partial x}.
$$

Note, that since equation (\ref{reg_part_0}) is quasi-linear, and equation (\ref{reg_part_1}) is linear, their solutions can be found, for example, by means of the method of characteristics. So, we can assume that these solutions are known.

Equations for the singular terms of the asymptotics are written as follows
\begin{equation} \label{singular_part_x_0}
\frac{\partial^2 V_0}{\partial \tau^2} + a_0(x, t) \, \varphi^{\,\prime}(t) \, \frac{\partial V_0}{\partial\tau} - b_0(x, t) \left( u_0
\frac{\partial V_0}{\partial\tau} + V_0 \frac{\partial V_0} {\partial\tau} \right) = 0,
\end{equation}
\begin{equation} \label{singular_part_x_1}
\frac{\partial^2 V_j}{\partial \tau^2} + a_0(x, t) \, \varphi^{\,\prime}(t) \, \frac{\partial V_j} {\partial\tau} - b_0(x, t) \left( u_0
\frac{\partial V_j} {\partial\tau} + \frac{\partial }{\partial \tau} \left( V_0 V_j \right) \right) = {F}_j (x, t, \tau),
\end{equation}
and found after substituting solution ({\ref{as_sol}}) into (\ref{Burgers_v}) and taking into account equations (\ref{reg_part_0}) and
(\ref{reg_part_1}).

Here the functions
$$
{F}_j (x,t, \tau)=F_j(t, V_0(x, t, \tau), \ldots , V_{j-1} (x, t, \tau), u_0(x, t), \ldots ,  u_j(x, t)),
$$
where $ j=1,2, \ldots \, $, are found recurrently after finding the regular part of the asymptotics.

Unlike the equations for the regular part of the asymptotics, the equations (\ref{singular_part_x_0}), (\ref{singular_part_x_1}) generally cannot be integrated in a closed form
for $ j=0,1, \ldots \, $, because these equations have variable coefficients and their solutions have to belong to the functional space $G$.
Therefore, let us consider the problem of existence of  these solutions.

The study is done as follows:

1. we assume that the curve $ \Gamma $ is known and equations (\ref{singular_part_x_0}), (\ref{singular_part_x_1}) are considered on the curve $ \Gamma $. The equation for the main term is integrated by quadratures, and for the other equations it is possible to lower their order. As a result, the first order linear inhomogeneous differential equations are deduced, solutions of which are written explicitly. Then the problem of belonging found solutions to the space $ G $ is studied;

2. from the condition that solutions to equations (\ref{singular_part_x_0}), (\ref{singular_part_x_1}) belong to the space $ G $ on the curve $ \Gamma $, we obtain non-linear ordinary differential equation for the function $ \varphi = \varphi(t) $.

Further, we use notation $ v_j = v_j(t, \tau) = V_j (x, t, \tau)\bigr|_{x = \varphi(t)}, $ $ j = 0, 1, \ldots \, $.

\subsection{The main term of the singular part of the asymptotics}

The function $ v_0(t, \tau) = V_0(\varphi (t), t, \tau )$ satisfies the following nonlinear partial differential equation
\begin{equation} \label{sing_part_0}
\frac{\partial^2 v_0}{\partial \tau^2} + a_0(\varphi, t) \, \varphi^{\,\prime}(t) \, \frac{\partial v_0}{\partial\tau} - b_0(\varphi, t) \left( u_0(\varphi, t)  + v_0 \right) \, \frac{\partial v_0}{\partial\tau} = 0.
\end{equation}

Since the function $ V_0(x, t, \tau) $ belongs to the space $ G $, we find the function $ v_0( t, \tau) $ as
\begin{equation} \label{sol_sing_0}
v_0(t, \tau) = A \, [1-\tanh{\left( \beta \, \, (\tau + c ) \right) } ] ,
\end{equation}
where
\begin{equation} \label{form_A}
A = A (t) = \frac{a_0(\varphi(t), t) \varphi^{\,\prime}(t) - b_0(\varphi(t), t)u_0(\varphi(t), t)}{b_0(\varphi(t), t)} ,
\end{equation}
\begin{equation} \label{form_beta}
\beta = \beta (t) = \frac{1}{2} \, A(t) \, b_0(\varphi(t), t),
\end{equation}
$ c=c(t) $ is a \textit{constant} of integration, which is assumed to be zero.

Next we show that as the main term of the singular part of the asymptotics $ V_0(x, t, \tau) $  we can consider the function $ v_0( t, \tau) $ that is given by formula  (\ref{sol_sing_0}).

Considering the main term of the regular part of the asymptotics is known, we prove the following theorem.

{\bf Theorem 1.} {\it Let condition $a_0(x,t) \,  b_0(x,t) \not=0 $ be fulfilled for all $ (x,t) \in {\bf R} \times [0; T] $, where the functions $a_0(x,t)$, $ b_0(x,t) $ are infinitely differentiable in the variables $ (x,t) \in {\bf R} \times [0; T] $. Then for any infinitely differentiable function $ x = \varphi(t) $, $ t \in [0; T] $, the following function
\begin{equation} \label{as_sol_0}
Y_0(x,t,\varepsilon) = u_0(x, t) + v_0(t, \tau), \quad \tau =
\frac{x-\varphi(t)}{\varepsilon},
\end{equation}
is the main term of the asymptotic step-like solution to equation (\ref{Burgers_v}) and satisfies the equation with an asymptotic accuracy $ O(1) $ on the set $ {\mathbf R}
\times[0; T] $. Moreover, as $ \tau \to + \infty $ it equates (\ref{Burgers_v}) with an asymptotic accuracy $ O(\varepsilon) $. 

If additionally for function (\ref{form_A}) the condition $ \displaystyle \frac{dA}{dt} = 0 $ holds then (\ref{as_sol_0}) satisfies equation (\ref{Burgers_v}) with an asymptotic accuracy $ O(\varepsilon) $ as $ \tau\to - \infty $. }

{\bf Proof.} For proving theorem 1 it is necessary to estimate the residue for (\ref{as_sol_0}) in the case of equation (\ref{Burgers_v}). We find an explicit expression for the residual function by substituting asymptotic solution (\ref{as_sol_0}) into equation  (\ref{Burgers_v}) and taking into account the equations for the regular and the singular parts of the asymptotics.

The searched residual function is determined by formula
\begin{equation} \label{nevyazka_sol_0}
g_0(x,t,\varepsilon) = \frac{1}{\varepsilon} \left[ a_0(x,t) - a_0(\varphi(t), t) \right] \varphi^{\, \prime}(t) \frac{\partial v_0}{\partial\tau} +
\end{equation}
$$
+ \frac{1}{\varepsilon} \left[ u_0(x,t)b_0(x, t) - u_0(\varphi(t), t) b_0(\varphi(t), t) \right] \varphi^{\, \prime}(t) \frac{\partial v_0}{\partial\tau} +
$$
$$
+ \frac{1}{\varepsilon} \left[ b_0(x,t) - b_0(\varphi(t), t) \right] v_0 \frac{\partial v_0}{\partial\tau} + a_0(x, t) \frac{\partial
v_0}{\partial t} + a_1(x, t) \frac{\partial v_0}{\partial \tau} +
$$
$$
+ b_1(x, t) \left(u_0 + v_0 \right)\frac{\partial v_0}{\partial \tau} + O(\varepsilon).
$$


For all $ (x,t) \in \Omega_\mu (\Gamma ) = \{(x; t) \in {\mathbf R}\times[0; T]: |x - \varphi(t)| < \mu,\} $, where $ \mu > 0 $, we have inequalities
$$
|a_0(x, t) - a_0(\varphi(t), t)| \le C_1 |x - \varphi(t)|, 
$$
$$
|b_0(x, t) - b_0(\varphi(t), t)| \le C_2 |x - \varphi(t)|,
$$
$$
|u_0(x, t) b_0(x, t) - u_0(\varphi(t), t) b_0(\varphi(t), t)| \le C_3 |x - \varphi(t)|
$$
with some positive constants $ C_k $, $ k = \overline{1, 3}. $

Because $ \displaystyle \frac{\partial v_0}{\partial\tau} $ is quickly decreasing function with respect to the variable $ \tau $, the asymptotic equality \ $ g_0(x, t, \varepsilon) = O(1) $ is true.

Since $ v_0(t, \tau) \in G $, the function $ v_0(t, \tau) $ obviously satisfies equation (\ref{Burgers_v}) with accuracy $ O(\varepsilon) $ as $ \tau \to +\infty $.
If the value of $ A(t) $ in (\ref{form_A}) is a constant, then the function $ v_0(t, \tau) \in G $ satisfies condition $ v_{0t}(t, \tau) \in G_0 $. Thus, the function $ v_0(t, \tau) $ satisfies equation (\ref{Burgers_v}) with an asymptotic accuracy $ O(\varepsilon)$ as $ \tau \to -\infty $.

Theorem 1 is proven.

\subsection{Higher terms of the singular part of the asymptotics}

Let us consider equations for higher terms of the singular part of the asymptotics on the discontinuity curve $ \Gamma $. Remind that the functions  $ v_j = v_j(t, \tau), $ $ j \in {\mathbf N} $, belong to the space $ G $. We have
\begin{equation} \label{sing_part_j}
\frac{\partial^2 v_j}{\partial \tau^2} + a_0(\varphi, t) \varphi^{\,\prime}(t) \, \frac{\partial v_j}{\partial\tau} - b_0(\varphi, t) \left[ u_0(\varphi, t) \frac{\partial
v_j}{\partial\tau} + \frac{\partial}{\partial\tau} \left( v_0 v_j \right) \right] = {\cal F}_j (t, \tau),
\end{equation}
where $ {\cal F}_j (t, \tau) = F_j(t, V_0(x, t, \tau), \ldots , V_{j-1} (x, t, \tau), u_0(x, t), \ldots , u_j(x, t)) \bigr|_{x = \varphi(t)}, $ are defined recurrently after finding the functions $u_0(x,t)$, $ u_1(x,t) $, $\ldots $, $ u_j(x,t)$, $ V_0(x,t, \tau), $ $ V_1(x, t, \tau) $, $ \ldots $, $ V_{j-1} (x,t, \tau)$, $j=\overline{1,N}$.

In particular,
\begin{equation} \label{form_cal_F}
{\cal F}_1(t, \tau) = a_0(\varphi, t) \frac{\partial v_0}{\partial t} + b_0(\varphi, t) u_{0 x}(\varphi, t) v_0 +
\end{equation}
$$
+ \left[-a_1(\varphi, t) \varphi^{\,\prime}(t) + b_1(\varphi, t) u_0(\varphi(t), t) + b_0(\varphi, t) u_1(\varphi, t)\right]\frac{\partial v_0}{\partial \tau}+
$$
$$
+ \tau\left[ -a_{0 x}(\varphi, t) \varphi^{\,\prime}(t) + b_{0 x}(\varphi, t) u_0(\varphi, t) + b_0(\varphi, t) u_{0 x}(\varphi, t)\right] \frac{\partial v_0}{\partial \tau} +
$$
$$
+ \left[ b_1(\varphi, t) + \tau b_{0 x}(\varphi, t) \right] v_0 \frac{\partial v_0}{\partial \tau } .
$$

Since the variable $ t $ can be considered as a parameter in the equation, we integrate this equation in the variable $ \tau $. We obtain
\begin{equation} \label{sol_sing_part_j}
\frac{\partial v_j}{\partial\tau} = \left[ -a_0(\varphi(t), t) \varphi^{\,\prime}(t) + b_0(\varphi(t), t) \left( u_0(\varphi(t), t) + v_0 \right) \right] v_j + \Phi_j(t, \tau),
\end{equation}
where
\begin{equation} \label{formula_Phi_j}
\Phi_j(t, \tau) = \int\limits_{\tau_0}^\tau {\cal F}_j(t, \xi) d \xi, \quad j \in {\mathbf N}.
\end{equation}

Solution to equation (\ref{sol_sing_part_j}) can be explicitly written in the following form
\begin{equation} \label{sol_sing_part_j_1}
v_j(t, \tau) = \left( C_0 \, \cosh^{2}(\beta\tau_0) + \int\limits_{\tau_0}^\tau \cosh^{2}(\beta s) \Phi_j(t, s) d s \right) \cosh^{-2}(\beta\tau),
\end{equation}
where $ C_0 $, $ \tau_0 $ are arbitrary constants and the value $ \beta = \beta (t) $ is defined with formula (\ref{form_beta}), $ j \in {\mathbf N} $.

The following lemma is true.

{\bf Lemma 1.} {\it If $ \Phi_j(t, \tau) \in G $, $ j \in {\mathbf
N} $, then for any $ t \in [0;T] $ function (\ref{sol_sing_part_j_1}) is a bounded solution of equation
(\ref{sol_sing_part_j}).}

Statement of lemma 1 concludes from the following inequality
$$
\left| \cosh^{-2}(\beta\tau) \int\limits_{\tau_0}^\tau
\cosh^{2}(\beta s) \Phi_j(t, s) d s \right| \le
$$
$$
\le c_{1j} \,
\cosh^{-2}(\beta\tau) \, \left(\frac{\sinh{(2\beta \tau)}}{4\beta} +
\frac{\tau}{2} + c_{2j} \right) <  \infty ,
$$
where $ c_{1j} $, $ c_{2j} $ are some constants, $ j \in {\mathbf N} $.

Lemma 1 is proven.

Let us consider function (\ref{sol_sing_part_j_1}). It is easy to notice such its properties:

1. the function $ v_0(t, \tau) $ is written via \, $ \tanh{( \beta \, \tau ) } $;

2. the right-side function (\ref{sing_part_j}) depends on $ v_k(t,
\tau) $, 
derivatives of the functions with respect to the variables $\tau $ and their products.

So, it is possible to assume that under some conditions on the coefficients of series (\ref{coeff}) the function $ \Phi_j(t, \tau) $, $ j \in {\mathbf N}$, may have the following form
\begin{equation} \label{form_Phi}
\Phi_j(t, \tau) = C_{10}(t) + \sum\limits_{k=1}^3 C_{1k}(t)
\tanh^{k}(\beta \tau) + \sum\limits_{k=1}^m C_{2k}(t)
\cosh^{-k}(\beta \tau),
\end{equation}
where $ C_{1k}(t), k=\overline{0,3} $, $ C_{2k}(t) $, $ k=\overline{1,m} $, are some values, $ m $ is a natural, $ t $ is a parameter.

It is easy to see, that, under some conditions on the coefficients of expansion in (\ref{form_Phi}), function (\ref{sol_sing_part_j_1}) belongs to the space
$ G $. Really, for example, let us study the first term of the singular part of the asymptotics on the curve $ \Gamma $ for the case of zero background, when the regular part of the asymptotics is trivial $ U_N(x,t,\varepsilon) = 0 $. Then we have
\begin{equation} \label{form_Phi_1}
\Phi_1(t, \tau) = \alpha_0 + \alpha_1 \tau + \alpha_2 \tau
\tanh{(\beta\tau)} + \alpha_3 \tau \tanh^2{(\beta\tau)} +
\end{equation}
$$
+ \alpha_4 \ln{\cosh{(\beta\tau)}} + \alpha_5 \tanh{(\beta\tau)}  + \alpha_6 \tanh^2{(\beta\tau)},
$$
where
\begin{equation} \label{alpha_0}
\alpha_0 = \alpha_0 (t) = - A \, a_1(\varphi(t), t) \, \varphi^{\,\prime}(t) + \frac{A^2}{2} \, b_1(\varphi(t), t),
\end{equation}
\begin{equation} \label{alpha_1}
\alpha_1 = \alpha_1 (t) = a_0(\varphi(t), t) \frac{d A}{d t}  - \frac{A^2}{2}\, b_{0x}(\varphi(t), t),
\end{equation}
\begin{equation} \label{alpha_2}
\alpha_2 =  \alpha_2 (t) = - \frac{A}{\beta} \, a_0(\varphi(t), t) \, \frac{d \beta}{d t} + A \, a_{0x}(\varphi(t), t) \, \varphi^{\,\prime} (t) - A^2 \, b_{0x}(\varphi(t), t),
\end{equation}
\begin{equation} \label{alpha_3}
\alpha_3 = \alpha_3 (t) = \frac{A^2}{2} \, b_{0x}(\varphi(t), t),
\end{equation}
\begin{equation} \label{alpha_4}
\alpha_4 = \alpha_4 (t) = - a_0(\varphi(t), t) \, \frac{d}{d t} \left( \frac{A}{\beta} \right) - \frac{A}{\beta} \,
a_{0x}(\varphi(t), t) \, \varphi^{\,\prime} (t) + \frac{A^2}{\beta} \, b_{0x}(\varphi(t), t) ,
\end{equation}
\begin{equation} \label{alpha_5}
\alpha_5 = \alpha_5 (t) = A \, a_1(\varphi(t), t) \, \varphi^{\,\prime} (t) - A^2 b_1(\varphi(t), t) + \frac{A^2}{2 \beta} \, b_{0x}(\varphi(t), t),
\end{equation}
\begin{equation} \label{alpha_6}
\alpha_6 = \alpha_6 (t) = \frac{A^2}{2} \, b_1(\varphi(t), t) .
\end{equation}
Here the values $ A $, $ \beta $ are defined by formulas (\ref{form_A}), (\ref{form_beta}).

Since the function $ \Phi_1(t, \tau) $ must be bounded as an element of the space $ G $, it is necessary the following condition 
\begin{equation} \label{cond_Phi}
\alpha_1 = \alpha_2 = \alpha_3 = \alpha_4 = 0
\end{equation}
takes place.

Using formulas (\ref{alpha_1}) -- (\ref{alpha_4}) we write condition (\ref{cond_Phi}) in the terms of the coefficients of equation (\ref{Burgers_v}).
From condition $ \alpha_3 = 0 $ it is deduced equality $ b_{0x}(\varphi (t), t) = 0 $, that is fulfilled, for example, in the case $ b_{0}(x,t) = b_{0}(t) $. In this case from condition $ \alpha_1 = 0 $ we find relation $ A^{\, \prime}(t) = 0 $, that is equivalent to non-linear second order ordinary differential equation for the discontinuity curve
\begin{equation} \label{eq_rozryv}
\frac{d\varphi}{d t}  = \rho \, \frac{b_0(t)}{a_0(\varphi, t)} ,
\end{equation}
where $ \rho \not= 0$ is an arbitrary constant.

Here the function $ \varphi = \varphi (t) $ is a solution to equation (\ref{eq_rozryv}) with initial condition $ \varphi (0) = \varphi_0 $, defined on some (maximal)
interval $ (\omega_-, \omega_+ ) \subset {\mathbf R} $, where the values  $ \omega_-, \omega_+ $ depend on initial value $ \varphi_0 $. In general case the interval $ (\omega_-, \omega_+ ) $ is finite, although, above all, of considerable interest is the case when the function $ \varphi = \varphi (t) $ is defined  for all $ t \ge 0 $.

Under condition (\ref{eq_rozryv}) equality $ \alpha_2 = \alpha_4 =0 $ is equivalent to relation
\begin{equation} \label{con_a0,b0}
a_0 ( \varphi (t), t) \, b^{\, \prime}_0 (t) = \rho \, b^{\, 2}_0 (t) \,
a_{0x}( \varphi (t), t).
\end{equation}

Thus, if the coefficients $ a_0(x, t) $, $ b_0(x, t) $ in (\ref{coeff}) satisfy conditions $ a_0(x, t) = a_0(t), $ $ b_0(x, t) = b_0 = const $, and $ \varphi = \varphi(t) $ is a solution to equation (\ref{eq_rozryv}), then the function $ \Phi_1(t, \tau) $ belongs to the space $ G $. In addition, according to lemma 1 the solution $ v_1(t, \tau) $
of equation (\ref{sol_sing_part_j}) is a bounded function that can be analytically obtained from formula (\ref{sol_sing_part_j_1}).

\subsubsection{Asymptotic estimate for the first approximation}

Let $ \tau_0 = 0 $ in (\ref{sol_sing_part_j_1}) and conditions (\ref{cond_Phi}) be fulfilled.
Taking into account representation (\ref{form_Phi_1}), we find the function $ v_1(t, \tau) $ explicitly. We have
\begin{equation} \label{form_v_1}
v_1(t, \tau) = \frac{\alpha_5}{2 \beta} + \left[ c - \frac{\alpha_5}{2 \beta} + ( \alpha_0 - \alpha_6) \, \frac{\tau}{2} \, \right] \cosh^{-2}{(\beta\tau)} + \frac{\alpha_0 + \alpha_6}{2 \beta} \, \tanh{(\beta\tau)},
\end{equation}
where the values $ \alpha_0 $, $ \alpha_5 $, $ \alpha_6 $ are defined by (\ref{alpha_0}), (\ref{alpha_5}), (\ref{alpha_6}), and $ c $ is a constant of integration.

Obviously the constructed function $ v_1(t, \tau) \in G $. Then the first asymptotic approximation for solution of equation (\ref{Burgers_v}) is written as
\begin{equation} \label{as_sol_1}
Y_1(x, t, \varepsilon) = v_0(t, \tau) + \varepsilon v_1(t, \tau), \quad \tau = \frac{x-\varphi(t)}{\varepsilon},
\end{equation}
where $ v_0(t, \tau) $, $ v_1(t, \tau) $ are defined by formulas (\ref{sol_sing_0}), (\ref{form_v_1}) respectively.

The property is deduced from the following theorem.

{\bf Theorem 2.} { \it Let the following conditions be fulfilled:

$ 1^0 $. the functions $a_0(x,t)$, $ b_0(x,t) $, $a_1(x,t)$, $ b_1(x,t) $ are infinitely differentiable with respect to the variables $ (x,t) \in {\bf R} \times
[0; T] $ \, and \, $a_0(x,t)\not=0$, $ b_0(x,t)=b_0(t) \not=0 $;

$ 2^0 $. the function $ \varphi = \varphi (t) $, $ t \in [0;T] $, is a solution of the  differential equation  (\ref{eq_rozryv});

$ 3^0 $. relation (\ref{con_a0,b0}) takes place.

Then function (\ref{as_sol_1}) is the first approximation for the asymptotic step-like solution for equation  (\ref{Burgers_v}) and satisfies equation (\ref{Burgers_v}) with accuracy $ O(\varepsilon) $ on the set $ {\mathbf R} \times[0; T] $, and satisfies this equation with accuracy $ O(\varepsilon^2) $ for $ \tau\to +\infty $.

If additionally condition
\begin{equation} \label{cond_v_1}
\frac{d}{dt} \, \frac{ a_1(\varphi(t), t) \varphi'(t) - b_1(\varphi(t), t) A(t)}{ b_0(\varphi (t), t) } = 0
\end{equation}
holds, where the value $ A (t) $ is given with formula (\ref{form_A}), then function (\ref{as_sol_1}) satisfies equation (\ref{Burgers_v}) with an asymptotic accuracy $ O(\varepsilon^2) $ as $ \tau\to - \infty $. }

{\bf Proof.} As in the case of proving theorem 1 it is necessary to estimate residual value for function (\ref{as_sol_1}) and equation (\ref{Burgers_v}).
Substituting asymptotic solution (\ref{as_sol_1}) to equation (\ref{Burgers_v}) and taking into account the equations for the singular terms we get
explicit expression for the residual value. Thus, we have to estimate function
\begin{equation} \label{nevyazka_sol_1}
g_1(x, t, \varepsilon) = \left[a_0(x,t) - a_0(\varphi(t), t) \right] \frac{\partial v_0}{\partial t} -
\end{equation}
$$
- \left[\frac{1}{\varepsilon}\left(a_0(x, t) - a_0(\varphi(t), t) - \varepsilon \tau a_{0x}(\varphi(t), t) \right) + \left(a_1(x, t) - a_1(\varphi(t), t) \right)\right] \times $$
$$
\times \varphi' \, \frac{\partial v_0}{\partial \tau} 
- \left[a_0(x,t) - a_0(\varphi(t), t) \right] \varphi' \, \frac{\partial v_1}{\partial \tau}+
$$
$$
+ \varepsilon \left[a_1(x, t) \frac{\partial v_0}{\partial t} - a_2(x, t) \varphi' \, \frac{\partial v_0}{\partial \tau} + a_0(x, t) \frac{\partial v_1}{\partial t} - a_1(x, t) \varphi' \, \frac{\partial v_1}{\partial \tau} \right] +
$$
$$
+ \left[\frac{1}{\varepsilon}\left(b_0(x, t) - b_0(\varphi(t), t) - \varepsilon \tau b_{0x}(\varphi(t), t) \right) + \left(b_1(x, t) - b_1(\varphi(t), t) \right)\right] \times
$$
$$
\times  v_0 \, \frac{\partial v_0}{\partial \tau} 
+ \left[ b_0(x, t) - b_0(\varphi(t), t) \right] \left( v_0 \, \frac{\partial v_1}{\partial \tau} +  v_1 \, \frac{\partial v_0}{\partial \tau}\right)+
$$
$$
+ \varepsilon \left[b_1(x, t) \left( v_0 \, \frac{\partial v_1}{\partial \tau} +  v_1 \, \frac{\partial v_0}{\partial \tau}\right)+ b_2(x, t) v_0 \, \frac{\partial v_0}{\partial \tau}  +  b_0(x, t) v_1 \, \frac{\partial v_1}{\partial \tau}\right] + O\left(\varepsilon^2 \right).
$$

Consider the first two terms of the residual function (\ref{nevyazka_sol_1}).
Since the explicit formula for $ v_0 (t, \tau) $ and condition $ 2^0 $ directly deduce $ v_{0t} \in G_0 $, for the first term we obtain the asymptotic estimate
$$
\left| \left[a_0(x, t) - a_0(\varphi(t), t) \right] \frac{\partial v_0}{\partial t} \right| \le \left|a_0(x,t) -
a_0(\varphi(t), t) \right| \, \left| \frac{\partial v_0}{\partial t} \right|  \le
$$
$$
\le \, M_1 \varepsilon |\tau| \left| \frac{\partial v_0}{\partial t} \right| \le C_1 \varepsilon,
$$
where $ M_1 > 0 $, $ C_1 > 0 $ are certain values.

Analogously for the second term of the residual function (\ref{nevyazka_sol_1}) we have inequalities:
$$
\left| \frac{1}{\varepsilon} \left[a(x, t) - a_0(\varphi(t), t) - \varepsilon \, \tau \, a_{0x}(\varphi(t), t) \right] \, \frac{\partial v_0}{\partial \tau} \right| \le
$$
$$
\le \frac{1}{\varepsilon} \, M_2 \, \varepsilon^2 \, \tau^2 \, \left|\frac{\partial v_0}{\partial \tau} \right| \le C_2 \varepsilon,
$$
where $ M_2 > 0 $, $ C_2 > 0 $ are some constants.

Analogously the other terms of residual function (\ref{nevyazka_sol_1}) are estimated.

So, we have asymptotic equality \ $ g_1(x, t, \varepsilon) = O(\varepsilon). $

Because $ v_0(t, \tau), v_1(t, \tau) \in G $, the function $ Y_1(x, t, \varepsilon)$ obviously satisfies equation (\ref{Burgers_v}) with an asymptotic accuracy $ O(\varepsilon^2) $ as $ \tau \to +\infty $.

If additionally in formula (\ref{form_v_1}) value $ \displaystyle \frac{\alpha_0 + \alpha_6}{2\beta} $ doesn't depend on variable $t$, i.e. condition (\ref{cond_v_1}) holds, then the function $ Y_1(x, t, \varepsilon) \in G $ satisfies condition $ Y_{1t}(t, \tau) \in G_0 $. So, the function $ Y_1(x, t, \varepsilon) $ satisfies equation (\ref{Burgers_v}) with accuracy $ O(\varepsilon^2)$ as $ \tau \to -\infty $.

Theorem 2 is proven.

In this way, the higher terms of the singular part of the step-like asymptotics can be constructed, but these formulas are very cumbersome. Therefore they are not written here.
It should also be noted that for an adequate description of the solutions to the differential equations with a small perturbation, as a rule, it is sufficient to consider the asymptotic approximation of the first order \cite{Bogoliu_Mitropol, Naife, Zappale, Sam4}.

\section{Example} Let us consider the Burgers' equation with variable coefficients
\begin{equation} \label{Burgers_example}
\varepsilon u_{xx} = \left( t^2 + 1 + \varepsilon (x^2 + 1)^2  \right) u_t + \left(  1 + \varepsilon \, \frac{(x^2 + 1)^2}{t^2+1} \right) u u_x
\end{equation}
and construct the first approximation for its asymptotic solution for the case of zero background.

For this example the first coefficients of asymptotic series (\ref{coeff}), (\ref{as_sol}) are written as
\begin{equation} \label{Burgers_example_1}
a_0(x, t) = t^2 + 1, \quad a_1(x, t) = (x^2 + 1)^2,
\end{equation}
\begin{equation} \label{Burgers_example_2}
\ b_0(x, t) = 1, \quad b_1(x, t) = \frac{(x^2 + 1)^2}{t^2+1}.
\end{equation}

Taking into account (\ref{Burgers_example_1}), (\ref{Burgers_example_2}) and the equalities $ u_0(x, t)=u_1(x,t) = 0 $, from (\ref{eq_rozryv}) we get equations for the function
$ \varphi = \varphi (t) $ in the form: 
$$
(t^2+1) \varphi'(t) = \rho, \quad \rho \not= 0.
$$

Solution to this equation with initial condition $ \varphi(0) = 0 $ is written as
$$
\varphi(t) = \rho \arctan t
$$
and defined for all $ t \in {\mathbf R} $.

Then $ A=A(t)=\rho $, $ \beta = \rho /2 $ and you can easily make sure that conditions (\ref{con_a0,b0}) and (\ref{cond_v_1}) with $ C_0 = 0 $ are fulfilled.

Thus, all conditions of theorems 1, 2 are satisfied.

The main term of the singular part of the asymptotics for equation (\ref{Burgers_example}) is given with the following formula
\begin{equation} \label{Example_main_term}
V_0(x, t, \varepsilon) = 1 - \tanh{\left( \rho \, \, \frac{x - \rho \arctan t}{2\varepsilon} \, \right) } ,
\end{equation}
and the first term of the singular part of the asymptotics is written as
\begin{equation}\label{Example_first_term}
V_1(x, t, \varepsilon) = \left[ c - \rho^2 \ \frac{ \left( 1 + \rho^2 \arctan^2 t \right)^2}{t^2+1} 
\ \frac{x - \rho\arctan t}{2\varepsilon} \right] \times
\end{equation}
$$
\times \cosh^{-2} \, {\left( \rho \, \, \frac{x - \rho\arctan t}{2\varepsilon} \right)} .
$$

It is obvious that $ V_0(x, t, \varepsilon) \in G $, $ V_1(x, t, \varepsilon) \in G_0 $.

The function
\begin{equation}\label{Example_sol}
Y_1(x, t, \varepsilon) = 1 - \tanh \,  \left( \frac{x - \rho\arctan t}{2\varepsilon} \right)
+
\end{equation}
$$
+ \varepsilon \left[ c - \rho^2 \, \frac{\left( 1 + \rho^2 \arctan^2 t \right)^2}{t^2+1} \, \frac{x - \rho \arctan t}{2\varepsilon} \right] \,
\cosh^{-2} \, \left(\rho \, \frac{x - \rho\arctan t}{2\varepsilon}\right)
$$
is the first asymptotic step-like approximation for solution of equation (\ref{Burgers_example}) and according to theorem 2 satisfies the equation with accuracy
$ O(\varepsilon) $. Moreover, the approximation satisfies the equation with accuracy $ O(\varepsilon^2) $ for $ \tau \to \pm \infty $.

Graphs of functions (\ref{Example_main_term}), (\ref{Example_first_term}) and (\ref{Example_sol}) are demonstrated on plots 1 -- 6  for $ \rho = 1 $, $ c = 0 $, $ \varepsilon =0,9 $ and $ \varepsilon =0,25 $. These graphs demonstrate properties of solutions (\ref{Burgers_v}), that are inherent in functions and mathematically describe shock waves.

\begin{figure}[h]
\centering
\includegraphics[scale=0.7]{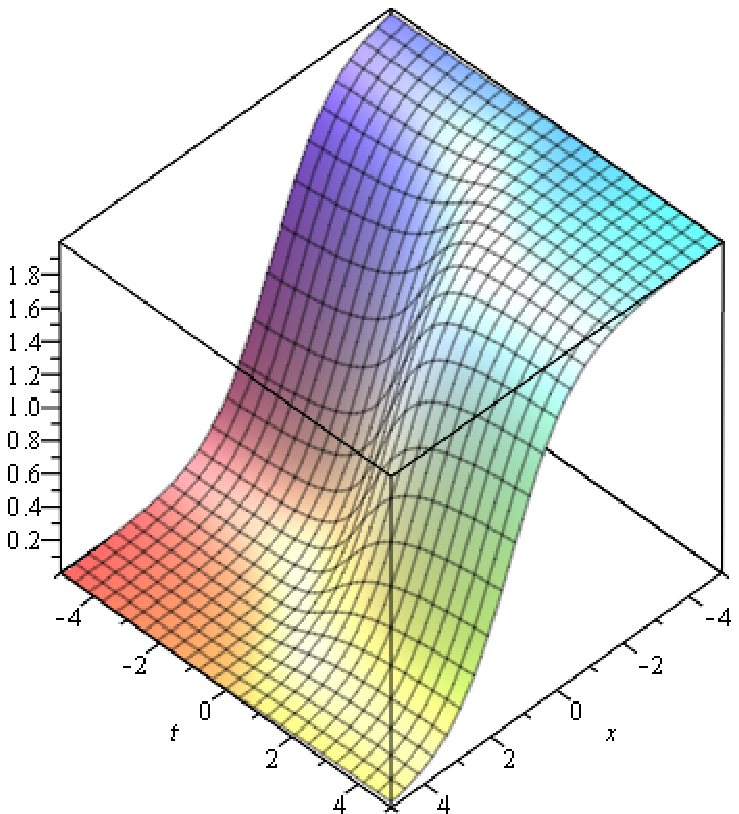}\qquad
\includegraphics[scale=0.7]{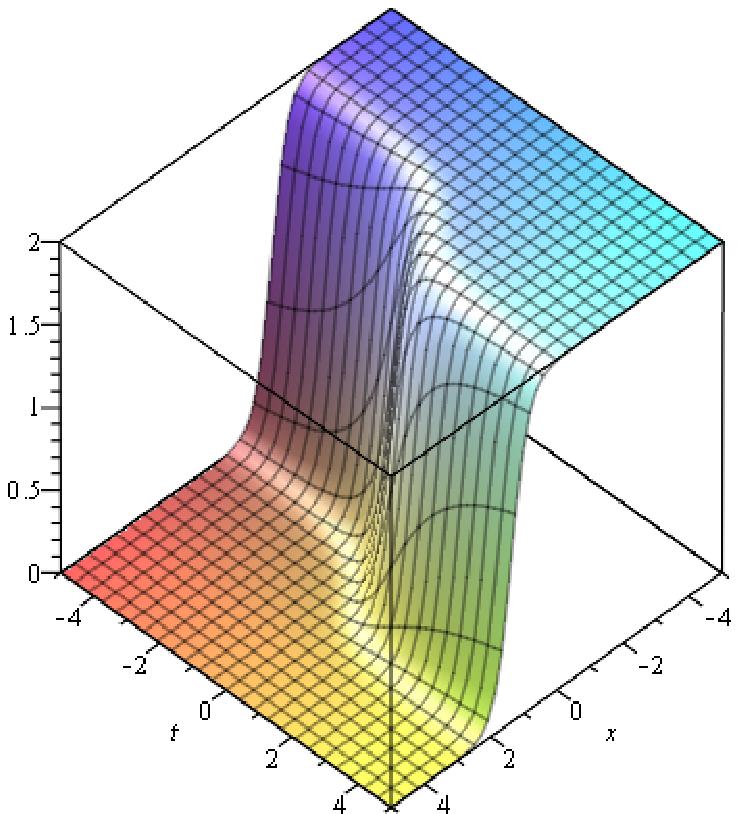}
\caption{ The main term of the asymptotic solution $V_0 (x, t,
\varepsilon)$ as $\varepsilon=0.9$ (at the left) and
$\varepsilon=0.25$ (at the right).} \label{fig:1}
\end{figure}

\begin{figure}[h]
\centering
\includegraphics[scale=0.7]{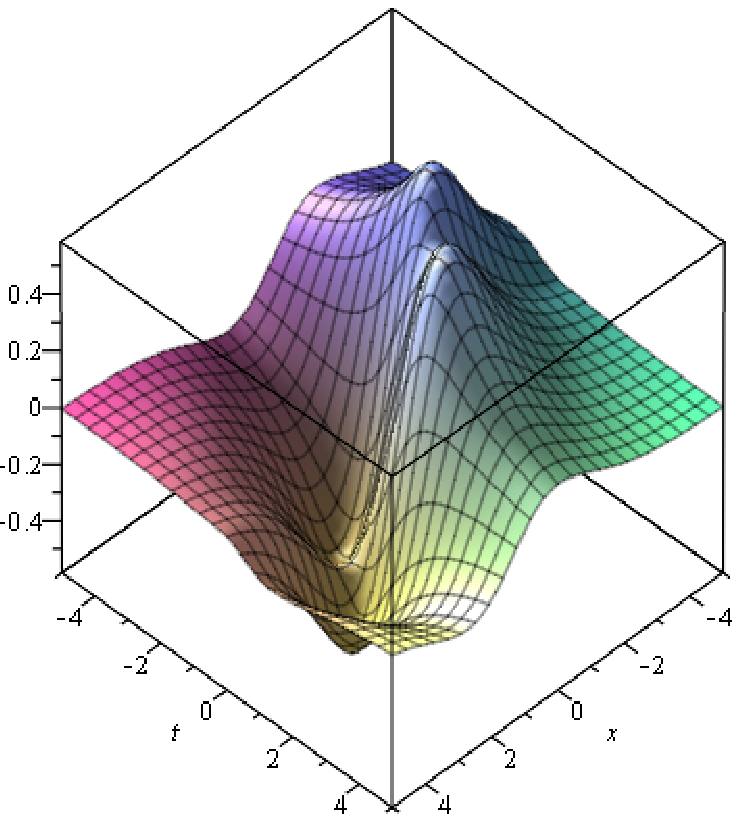}\qquad
\includegraphics[scale=0.7]{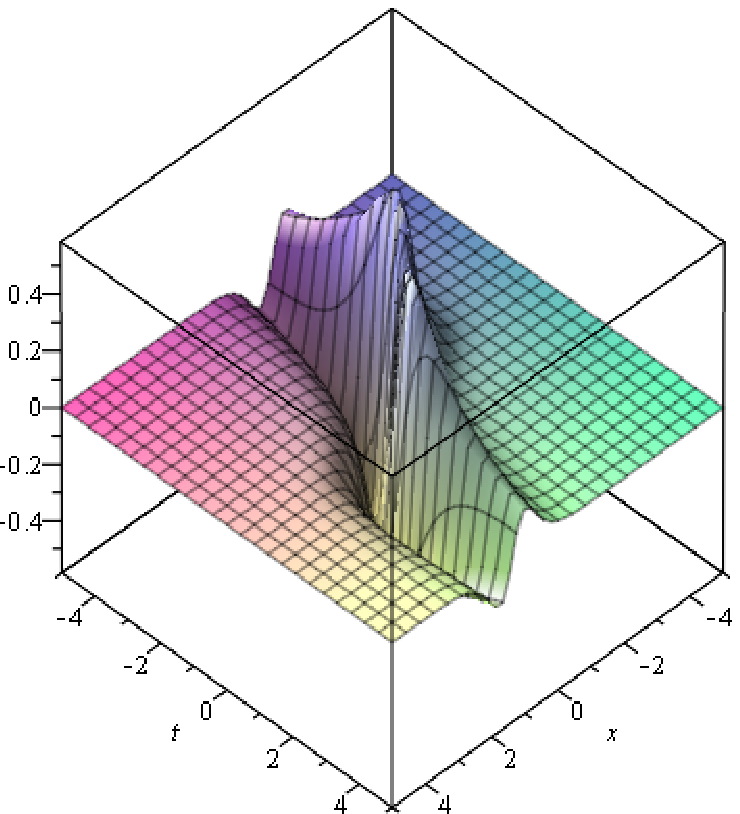}
\caption{ The first term of the asymptotic solution $V_1 (x, t,
\varepsilon)$ as $\varepsilon=0.9$ (at the left) and
$\varepsilon=0.25$ (at the right).} \label{fig:2}
\end{figure}

\begin{figure}[h]
\centering
\includegraphics[scale=0.7]{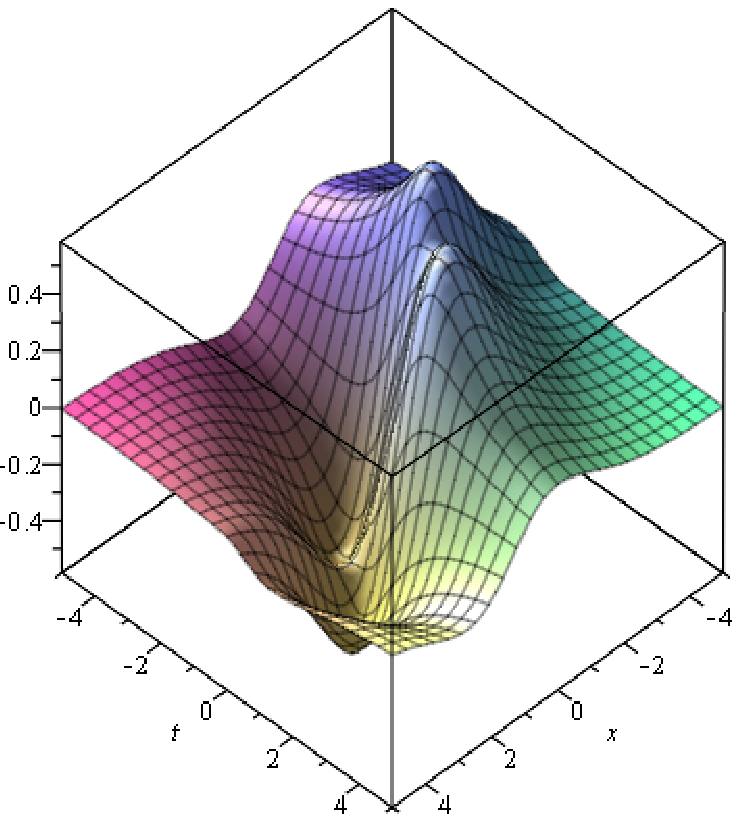}\qquad
\includegraphics[scale=0.7]{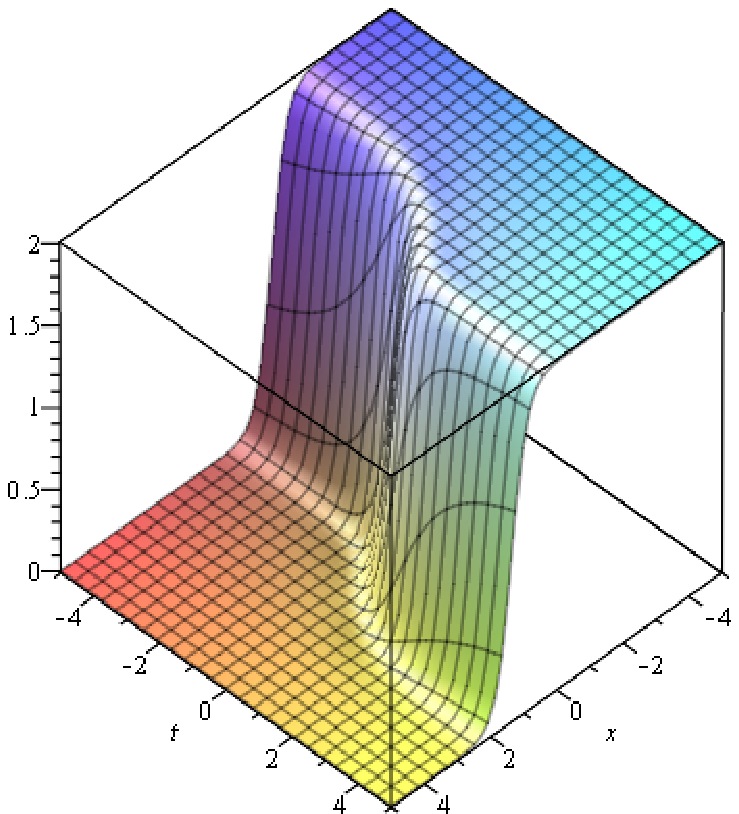}
\caption{ The first order asymptotic solution $Y_1 (x, t,
\varepsilon)$ as $\varepsilon=0.9$ (at the left) and
$\varepsilon=0.25$ (at the right).} \label{fig:3}
\end{figure}

\section{Conclusions and discussions}

The paper deals with the problem on constructing asymptotic step-like solutions to the Burgers' equation with variable coefficients and a small parameter. This equation can be considered as a mathematical model of wave processes in media with a small viscosity and variable characteristics and as generalization of classical Burgers' equation.

The Burgers' equation is known as the simplest mathematical model which unified typical non-linearity and viscosity and describes phenomena of hydrodynamical turbulence, blow-up and overturning (destruction) of waves. When the viscosity tends to zero, the limiting value of the solution is a discontinuous step-like function.

Naturally, the problem of finding asymptotic step-like solutions to the Burgers' equation with variable coefficients arises.
The problem is solved by means of the non-linear WKB method. In this paper, the algorithm of constructing these asymptotic solutions is proposed and statements on justification of the algorithm are proved. The obtained results are illustrated by an example, for which the first asymptotic step-like approximation is explicitly found.
There are also given graphs of these approximate solutions for certain numerical parameters.

In this example, the asymptotic solution, in contrast to the general case, is defined for all values of space and time variables, and has the form of the shock wave type function. These graphs also demonstrate that the first asymptotic approximation adequately describes solution of the equation under consideration.

\section{Acknowledgements}
Part of this work has been made when the first author was a visitor at Department of Basic and Applied Science for Engineering at Sapienza- University of Rome, through the program `Professori visitatori che si trovino in condizioni di rischio a causa di conflitti'. Hence he is very grateful for the support and the hospitality. The first and the last author also acknowledge the support of INdAM `Professori visitatori Sportello Ucraina'. The last author is a member of GNAMPA-INdAM, whose support is acknowledged.
The second author was supported by Programme PAUSE in Claude Bernard Lyon I University and she is very appreciative and grateful for the support
and the hospitality.


\end{document}